\documentclass[conference]{IEEEtran}
\IEEEoverridecommandlockouts
\usepackage{cite}
\usepackage{amsmath,amssymb,amsfonts}
\usepackage{algorithmic}
\usepackage{graphicx}
\usepackage{textcomp}
\usepackage{xcolor}
\usepackage{bm}
\usepackage{algorithm}
\usepackage{algorithmic}
\usepackage{multirow}

\newcommand\figref[1]{Fig.~\ref{#1}}
\newcommand\tabref[1]{Table~\ref{#1}}

\DeclareSymbolFont{lettersA}{U}{txmia}{m}{it}
\def\RRR{\mathbb{R}}

\newcommand{\argmax}{\mathop{\rm argmax}\limits}
\newcommand{\mean}{\mathop{\rm mean}\limits}
\newcommand{\median}{\mathop{\rm median}\limits}
\newcommand{\ctext}[1]{\raise0.2ex\hbox{\textcircled{\scriptsize{#1}}}}

\def\BibTeX{{\rm B\kern-.05em{\sc i\kern-.025em b}\kern-.08em
    T\kern-.1667em\lower.7ex\hbox{E}\kern-.125emX}}
\begin{document}

\title{Molecular activity prediction using graph convolutional deep neural network considering distance on a molecular graph}

\author{\IEEEauthorblockN{Masahito Ohue}
\IEEEauthorblockA{\textit{Dept. Computer Science,} \\
\textit{School of Computing,} \\
\textit{Tokyo Institute of Technology}\\
Tokyo, Japan\\
ohue@c.titech.ac.jp}
\and
\IEEEauthorblockN{Ryota Ii}
\IEEEauthorblockA{\textit{Dept. Computer Science,} \\
\textit{School of Computing,} \\
\textit{Tokyo Institute of Technology}\\
Tokyo, Japan\\
ii@bi.c.titech.ac.jp}
\and
\IEEEauthorblockN{Keisuke Yanagisawa}
\IEEEauthorblockA{\textit{Dept. Computer Science,} \\
\textit{School of Computing,} \\
\textit{Tokyo Institute of Technology}\\
Tokyo, Japan\\
yanagisawa@bi.c.titech.ac.jp}
\and
\IEEEauthorblockN{Yutaka Akiyama}
\IEEEauthorblockA{\textit{Dept. Computer Science,} \\
\textit{School of Computing,} \\
\textit{Tokyo Institute of Technology}\\
Tokyo, Japan\\
akiyama@c.titech.ac.jp}
}

\maketitle

\begin{abstract}
Machine learning is often used in virtual screening to find compounds that are pharmacologically active on a target protein. 
The weave module is a type of graph convolutional deep neural network that uses not only features focusing on atoms alone (atom features) but also features focusing on atom pairs (pair features); thus, it can consider information of nonadjacent atoms. However, the correlation between the distance on the graph and the three-dimensional coordinate distance is uncertain.
In this paper, we propose three improvements for modifying the weave module. First, the distances between ring atoms on the graph were modified to bring the distances on the graph closer to the coordinate distance.
Second, different weight matrices were used depending on the distance on the graph in the convolution layers of the pair features.
Finally, a weighted sum, by distance, was used when converting pair features to atom features.
The experimental results show that the performance of the proposed method is slightly better than that of the weave module, and the improvement in the distance representation might be useful for compound activity prediction.
\end{abstract}


\section{Introduction}
In drug research and development, it takes at least ten years to produce a single drug, and development costs are estimated to be several billion US dollars or more~\cite{new_drug}.
High-throughput screening methods for screening compounds that show activity against proteins targeted by drug discovery from large-scale compound libraries are popular~\cite{hts}; however, screening vast numbers of compounds is expensive. In contrast, virtual screening is expected to be able to predict active compounds (hit compounds) efficiently using a computer~\cite{cadd}.

One of the frameworks of virtual screening is a ligand-based method that uses machine learning to predict activity using known activity information as a teacher label~\cite{LBDD1, LBDD2}. In particular, in recent years, each atom of a compound is regarded as a node, and a bond is considered as an edge graph. Based on this, feature extraction can be performed using neural networks \cite{ngf, gcn, weave}. The graph convolutional neural network (GCN), which realizes the convolutional deep neural network by using a convolution operation on the graph structure, is used for such applications.

For graph feature extraction using GCN, neural graph fingerprints (NGF) \cite{ngf}, the GCN by Han {\it et al}. \cite{gcn} and the weave module \cite{weave} are often used.
These methods do not generate compound descriptors (feature vectors) based on a specific rule like ordinary fingerprints and have the advantage of being able to represent feature vectors by learning molecular structures flexibly.

NGF and Han's GCN do not consider edge features in the molecular graph but focus on learning the relationship with the nearest neighbor node.
On the other hand, the weave module of Kearnes {\it et al}. transforms feature vectors using pair features with distant atoms in addition to atom features focused only on atoms. Thus, the Weave module can consider features between distant atoms. However, the number of atoms forming a pair is different for each distance. Furthermore, the pair features of the Weave module cannot be considered in that respect.

In this paper, we propose a new improved GCN that can consider features between distant atoms by modifying the Weave module. In order to make effective use of the distance features on the molecular graph in the Weave module, we considered three improvements: correction of the distance on the molecular graph with respect to atoms in the ring structure, convolution method of pair features, and assembling of the pair features.

\section{Weave Module~\cite{weave}}
\begin{figure}[tb]
	\begin{center}
		\includegraphics[width=90mm]{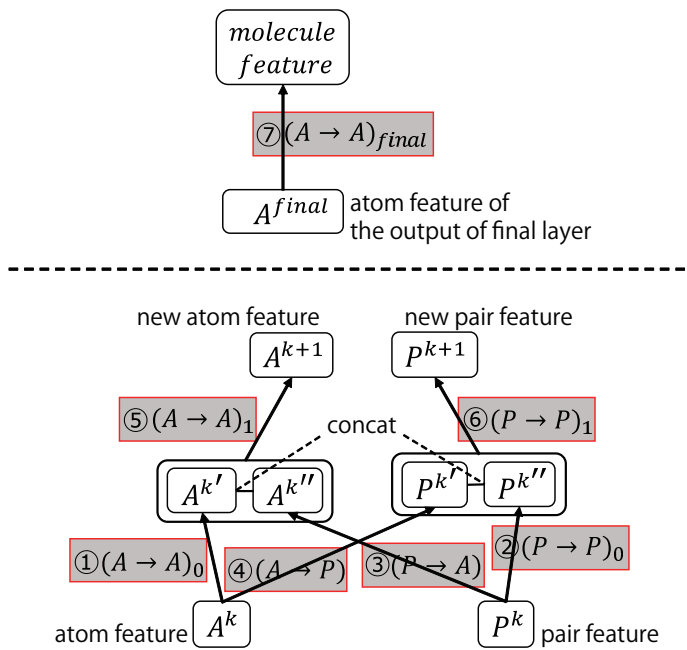}
		\caption{Weave module~\cite{weave}}
		\label{fig:weave_whole}
	\end{center}
\end{figure}

The network architecture of the Weave module \cite{weave} is shown in \figref{fig:weave_whole}. The Weave module consists of the seven transformations shown in \ctext{1}--\ctext{7} in~\figref{fig:weave_whole}.
This study was targeted at improving the method of generating the initial feature and transformation operation \ctext{3} (transforming from the pair feature to the intermediate atom feature). These operations are described as follows, and their further details can be found in ~\cite{weave}.

\subsection{Initial features}
Initial atom feature $\bm{A}^0 \in \RRR^{n_{\mathit{max}} \times d_a^0}$ and pair feature $\bm{P}^0 \in \RRR^{n_{\mathit{max}}^2 \times d_p^0}$, which are inputs to the network, use simple descriptors such as atom and bond types. $n_{\mathit{max}}$ is the maximum number of atoms in the molecule. $\bm{A}^0$ is a matrix in which $n_{\mathit{max}}$ number of $d_a^0$-dimensional feature vectors (row vectors) corresponding to one atom are vertically arranged.
$\bm{P}^0$ is a matrix in which $n_{\mathit{max}}^2$ number of $d_p^0$-dimensional feature vectors (row vectors) corresponding to one atom pair are vertically arranged.

\subsection{Transformation operation \ctext{3}: transform intermediate atom feature from pair feature}
\begin{figure}[tb]
	\begin{center}
		\includegraphics[width=90mm]{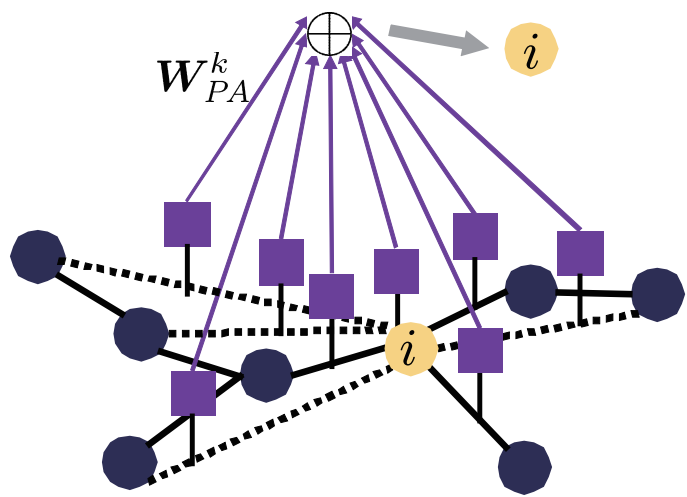}
		\caption{Converting pair features to atom features}
		\label{fig:weave_pa}
	\end{center}
\end{figure}

In Weave-module layer $k$, the following operation, as shown in \figref{fig:weave_pa} is performed on all the atom pairs comprising atom $i$. The intermediate atom feature for atom $i$ is calculated by adding them.
\begin{eqnarray}
\bm{a}_i^{k''}=\sum_{j}f\left(\bm{W}_{\mathit{PA}}^k \bm{p}_{(i,j)}^k+\bm{b}_{\mathit{PA}}^k\right)
\label{eq:weave_PA}
\end{eqnarray}
where $\bm{p}_{(i,j)}^k \in \RRR^{d_p^k}$ is an input pair feature vector of atom pair $(i,j)$ in the $k$-th layer,
$\bm{a}_i^{k''} \in \RRR^{d_{\mathit{PA}}}$ is an output atom feature vector of atom $i$,
$\bm{W}_{\mathit{PA}}^k \in \RRR^{d_{\mathit{PA}} \times d_p^k}$ is a weight matrix,
and $\bm{b}_{\mathit{PA}}^k \in \RRR^{d_{\mathit{PA}}}$ is a bias vector.
$f(\cdot)$ is an activation function that applies ReLU to all elements of a vector.
 
Atom feature $\bm{A}^{k''} \in \RRR^{n_{\mathit{max}} \times d_{\mathit{PA}}}$ is vertically arranged as $\bm{a}_i^{k''}$ for all atoms $i=1,...,n_{\mathit{max}}$.

\subsection{Point of issue}\label{chap:problem}
The following issues are present in the Weave module.

\begin{enumerate}
\item \textbf{Distance on the graph for atoms in a ring structure}\\
An uncertainty exists as to whether the distance on the graph and the real three-dimensional distance are correlated between atom pairs in the ring structure.

\item \textbf{Convolution of pair features}\\
Uniform weights are used for all pair features regardless of the distance length on the graph.

\item \textbf{Assembling of pair features}\\
All atoms in the pair are uniformly added to the convoluted pair feature, and the difference due to the distance between the pairs is not reflected.
\end{enumerate}

\section{Proposed Method}
Here, we introduce three improvements to solve the above-mentioned issues of the Weave module.

\subsection{Correction of distances related to atoms in ring structures (Prop. A)}
\begin{figure}[tb]
	\begin{center}
		\includegraphics[width=0.96\linewidth]{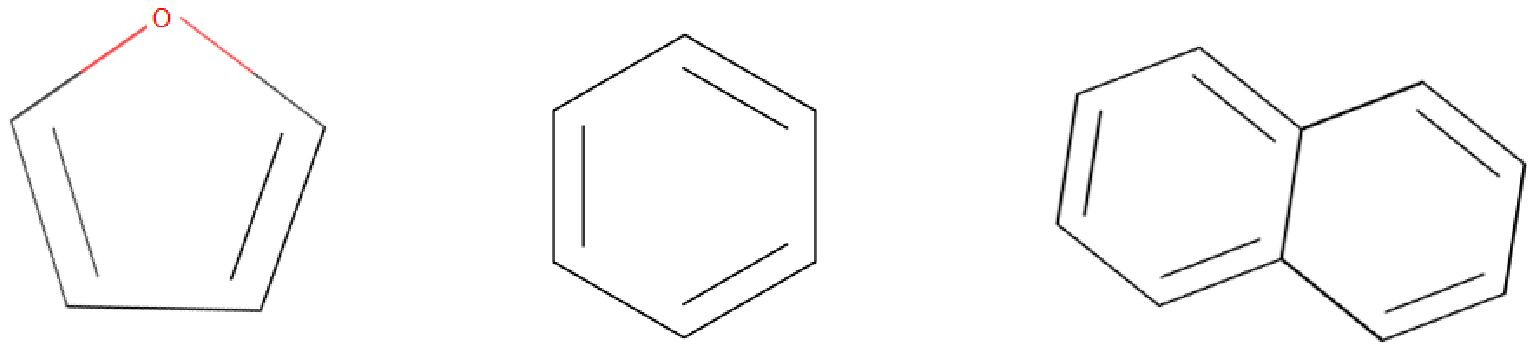}
		\caption{Examples of ring structure}
		\label{fig:ring_2d}
	\end{center}
	\begin{center}
		\includegraphics[width=0.92\linewidth]{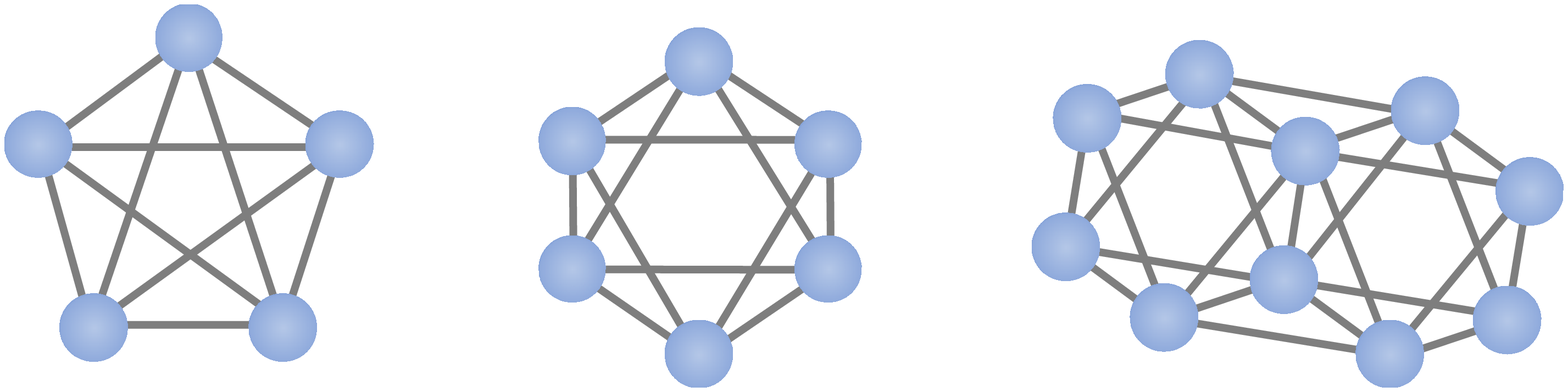}
		\caption{Examples of redefined ring structure (Prop. A)}
		\label{fig:ring_2d_def}
	\end{center}
\end{figure}

The pair feature of the Weave module defines the distance between atom pairs as the length of the shortest path on the graph.
The ring structure is relatively rigid in terms of the actual molecular conformation compared to the chain structure. Moreover, the distance on the conformation is shorter than the distance on the graph, considering two atoms in the molecule.
Therefore, with respect to the atom of interest, the atom pair at the orthoposition and metaposition is distance 1, and the atom pair at the para position is distance 2 (Figs. \ref{fig:ring_2d} and \ref{fig:ring_2d_def}).

To realize this definition of distance, we redefined the molecular graph by adding new edges to the atoms contained in the ring structure in the compound. The modified molecular graph is generated through {\bf Algorithm~\ref{algorithm1}}.

\begin{algorithm}[tb]                      
	\caption{Redefinition of distance on ring structure}
	\label{algorithm1}                          
	\begin{algorithmic}[1]                  
		\REQUIRE molecular graph $\mathcal{G}$
		\ENSURE redefined molecular graph $\mathcal{H}$
		\STATE $\mathcal{H} \leftarrow \mathcal{G}$
		\STATE $sssr \leftarrow \mathrm{GetSymmSSSR}(\mathcal{G})$
		\STATE $\mathit{ring\_atoms} \leftarrow \mathrm{flatten}(sssr)$
		\FOR{$ring$ in $sssr$}
		\FOR{$v_s$ in $ring$}
		\STATE $dict \leftarrow \mathrm{ShortestPathLength}(\mathcal{G}, v_s, \mathit{HOP})$
		\STATE $list \leftarrow [ \, ]$
		\FOR{$v_t, d$ in $dict.\mathrm{items}()$}
		\IF{$d=\mathit{HOP}$}
		\STATE $list.\mathrm{append}(v_t)$
		\ENDIF
		\ENDFOR
		\STATE $V_d \leftarrow \mathrm{set}(list) \cap \mathrm{set}(\mathit{ring\_atoms})$
		\FOR{$a$ in $V_d$}
		\STATE $\mathcal{H} \leftarrow \mathcal{H}\cap \mathrm{edge}(v_s, a)$
		\ENDFOR
		\ENDFOR
		\ENDFOR
	\end{algorithmic}
\end{algorithm}

In {\bf Algorithm~\ref{algorithm1}}, GetSymmSSSR() is used to obtain the symmetrized smallest set of smallest rings (SSSR) for a molecule, flatten() converts a nested list into a one-dimensional array, ShortestPathLength() obtains the shortest path length to a certain vertex $r$, and $\mathit{HOP}$ in the dictionary data type $dict$ is the number of hops from the focused atom to the atom at which the search is stopped.
Further, items() is used to obtain the searched atoms and the shortest path length from $dict$.
GetSymmSSSR() is implemented in RDKit (version 2018.03.4) \cite{rdkit} and ShortestPathLength() is implemented in NetworkX (version 2.2) \cite{nx}.

In this study, we defined $\mathit{HOP}=2$. In other words, the distance between atoms in the ring structure is regarded as $\left\lceil d/2 \right\rceil$. $d$ is the shortest path length between the atoms in a pair on the original graph. Taking furan (5-atom ring), benzene (6-atom ring), and naphthalene (polycyclic molecule) as examples, the graph redefined in this study is shown in \figref{fig:ring_2d}.

\subsection{Convolution of pair features with different weights (Prop. B)}
We improved the weights for pair features to be determined by learning the use of neural networks.
In the Weave module, pair features were convoluted using the same weight matrix, regardless of the distance length.
Therefore, we labeled distances $\mathit{dist_\mathrm{0}}, \mathit{dist_\mathrm{1}}, ..., \mathit{dist_n}, ..., \mathit{dist_{max}}, \mathit{dist_\infty}$ from the focus atom to distinguish each pair feature.
Here, $\mathit{dist_\infty}$ represents all distances greater than the maximum atomic pair distance $\mathit{dist_{max}}$.
We used different weight matrices $\bm{W}_{\mathit{PA}_{\mathit{dist_\mathrm{0}}}}, \bm{W}_{\mathit{PA}_{\mathit{dist_\mathrm{1}}}}, ..., \bm{W}_{\mathit{PA}_{\mathit{dist_n}}}, ..., \bm{W}_{\mathit{PA}_{\mathit{dist_{max}}}}, \bm{W}_{\mathit{PA}_\mathit{dist_\infty}}$ corresponding to these distances in the convolution of pair features.

\begin{figure}[tb]
	\begin{center}
		\includegraphics[width=0.96\linewidth]{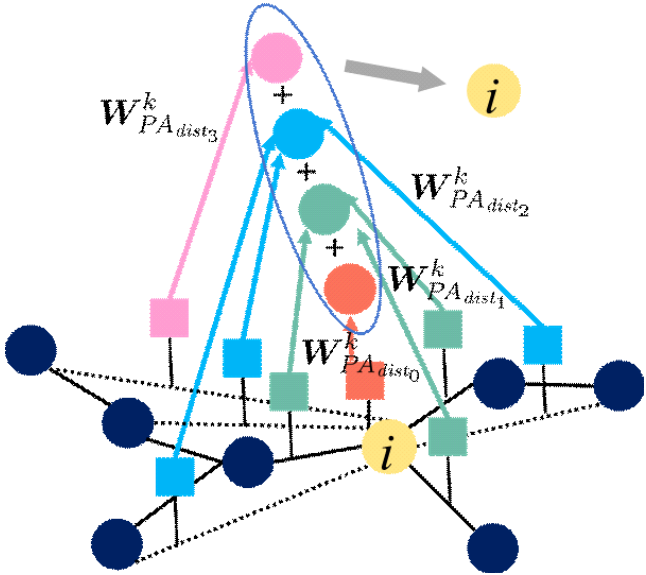}
		\caption{Convolution of pair features using different weights (Prop. B)}
		\label{fig:weave_weight}
	\end{center}
\end{figure}

In Prop. B, weight matrices according to the distance were used for atom pairs and convolution was performed. The intermediate atom feature of atom $i$ was calculated by taking the sum of atom pairs of atom $i$. This operation is shown in \figref{fig:weave_weight}.

\subsection{Assembling pair features based on distance (Prop. C)}
If the interatomic distance on the graph is large, the interatomic distance on conformation does not become constant. Atom pairs with large interatomic distances appear to be less important than those with small interatomic distances.
Therefore, when finding the intermediate atom feature $\bm{a}_i^{k''}$ of atom $i$, the closer the distance $d_{ij}$ is, the larger is the weighting performed by the three kinds of coefficients $g(d_{ij})$:
\begin{align}
g(d)&=0 \ \mathit{if} \ d>\mathit{dist_{max}} \ \mathit{else} \ 1 \ \ \mathrm{(step)}\\
g(d)&=-0.1d+1 \quad\quad\quad\quad\quad\quad\ \mathrm{(linear)}\\
g(d)&=1/d^2 \quad\quad\quad\quad\quad\quad\quad\quad\ \ \mathrm{(quadratic)}
\end{align}
This modifies Eq.~(1) as follows:
\begin{eqnarray}
\bm{a}_i^{k''}=\sum_{j}g(d_{ij})f\left(\bm{W}_{\mathit{PA}}^k \bm{p}_{(i,j)}^k+\bm{b}_{\mathit{PA}}^k\right)
\end{eqnarray}

\section{Experiments}

\subsection{Dataset}
We used the Biophysics datasets HIV, MUV, and PCBA from MoleculeNet~\cite{molnet}.
Molecular data are provided in SMILES format and converted to 2-D molecular graphs using RDKit~\cite{rdkit}. Hydrogen atoms were omitted, and compounds with the huge number of heavy atoms exceeding maximum number of atoms, $n_{\mathit{max}}$, were excluded from the dataset. The number of tasks in each dataset, number of active compounds, number of inactive compounds, number of compounds, and number of excluded compounds are shown in \tabref{table:dataset}. Given that the same compound is registered with different labels between each task, the numbers of active and inactive compounds were counted in duplicate.

\begin{table}[tb]
	\caption{Details of datasets}
	\label{table:dataset}
	\centering
	\begin{tabular}{lrrrrr}
		\hline
		\multicolumn{1}{c}{dataset} & \multicolumn{1}{c}{\#{}tasks} & \multicolumn{1}{c}{\#{}pos\footnotemark[1]} & \multicolumn{1}{c}{\#{}neg\footnotemark[1]} & \multicolumn{1}{c}{\#{}cmpds} & \multicolumn{1}{c}{\#{}excluded} \\ \hline
		HIV~\cite{hiv}                        & 1                        & 1,319                        & 39,065                         & 40,384                    & 743                   \\ 
		MUV~\cite{muv}                        & 17                       & 489                         & 249,397                        & 93,087                    & 0                     \\
		PCBA~\cite{pubchem}                       & 128                      & 471,273                      & 33,509,569                      & 437,035                   & 894                  \\\hline
	\end{tabular}
\begin{flushleft}
\footnotemark Given that the same compound is registered with different labels between each task, the number counted in duplicate as described.
\end{flushleft}
\end{table}

\subsection{Training and evaluation}
The GCN model was implemented using the deep learning library, Chainer Chemistry (version 0.4.0) \cite{chainer_chemistry}. 
The hyperparameters of GCN are listed in \tabref{table:param}. These were the same as those used by Kearnes {\it et al}. \cite{weave}.
We attempted to set maximum atom pair distance, $\mathit{dist_{max}}$, to 1--5.

In this study, the prediction performance of the model was evaluated using the ROC curve \cite{roc} and area under the curve (AUC), as shown in Eq.~(6), as well as the enrichment factor (EF) \cite{ef} in Eq.~(7) using the compound order arranged in descending order of prediction probability.
\begin{eqnarray}
	\mathrm{AUC}&=&1-\frac{1}{N_{\mathrm{Pos}}}\sum_{i=1}^{N_\mathrm{{Pos}}}\frac{N^i_\mathrm{{Neg}}}{N_\mathrm{{Neg}}}\\
	\mathrm{EF}_{x\%}&=&\frac{N_{\mathrm{Pos},x\%}/N_{x\%}}{N_{\mathrm{Pos}}/N},
\end{eqnarray}
where $N_\mathrm{{Pos}}$ is the number of active compounds, $N_{\mathrm{Neg}}$ is the number of inactive compounds, $N^i_{\mathrm{Neg}}$ is the number of inactive compounds ranked higher than the $i$-th active compound, $N$ is the number of compounds, $N_{\mathrm{Pos},x\%}$ is the number of active compounds in the top $x\%$, and $N_{x\%}$ is the number of $x\%$ compounds in the dataset (i.e., $N_{x\%}=\frac{x}{100}N$).

The AUC is 0.5 and 1.0 for random and complete predictions, respectively. $\mathrm{EF}_{x\%}$ is a value indicating how many times the active compound can be concentrated to the top $x\%$ through compound ranking, and EF = 1 in random prediction. In this study, $\mathrm{EF}_{1\%}$ and $\mathrm{EF}_{1\%}$ were used.

\begin{table}[tb]
	\caption{Model hyperparameters}
	\label{table:param}
	\centering
	\begin{tabular}{llr}
		\hline
		\multicolumn{2}{l}{hyperparameter}                                & value      \\ \hline
		\multicolumn{2}{l}{maximum number of atoms in molecule $n_{\mathit{max}}$}                          & 60        \\
		\multicolumn{2}{l}{maximum atomic pair distance $\mathit{dist_{max}}$}                          & 1--5       \\ 
		\multicolumn{2}{l}{Weave module\it{k}}                     & 2         \\ 
		\multicolumn{2}{l}{$d_{\mathit{AA}}, d_{\mathit{PP}}, d_{\mathit{PA}}, d_{\mathit{AP}}, d_A, d_P$}                               & 50        \\ 
		\multicolumn{2}{l}{$d_{A_{\mathit{final}}}$}                                & 128       \\ 
		\multicolumn{2}{l}{\#{}fully connected layers} & 2000, 100 \\ \hline
		\multirow{4}{*}{training}               & batch size        & 96        \\ 
		& optimizer     & Adam      \\ 
		& learning rate      & 0.001     \\ 
		& epoch         & 100       \\ \hline
		\multirow{2}{*}{train$:$valid$:$test}     & HIV     & $8:1:1$     \\  
		& PCBA, MUV     &$6:2:2$     \\ \hline
		\multirow{2}{*}{trial $m$} & HIV     & 10                     \\ 
		& PCBA, MUV     & 5      \\ \hline
	\end{tabular}
\end{table}

Each dataset was divided into the training data (train), validation data (valid), and test data (test) according to the ratio shown in \tabref{table:param}.
For each task in the dataset, we selected an epoch (learning checkpoint) that gives the best AUC for the validation data and applied it to the test data to calculate the averaged AUC value for each task. The AUC used in evaluation (AUC\_{}eval) is calculated as follows.
\begin{eqnarray}
	n_{{\mathit{best}},{\mathcal{T}}} &=& \argmax_{n} \mean_{i} \left(\mathrm{AUC}^{\mathrm{valid}}_{\mathcal{T},n,i}\right) \\
	\mathrm{AUC\_{}eval} &=& \median_{\mathcal{T}} \mean_{i} \left(\mathrm{AUC}^{\mathrm{test}}_{\mathcal{T},n_{{\mathit{best},\mathcal{T}}},i}\right),
	\label{eq:eval}
\end{eqnarray}
where $\mathcal{T}$ represents each task, $n$ is the epoch, and $i (= 1, ..., m)$ is the trial. 
$\mathrm{AUC}^{\mathrm{valid}}_{\mathcal{T},n,i}$ is an AUC value of task $\mathcal{T}$ of validation data using trained GCN with epoch $n$ in trial $i$.
$\mathrm{AUC}^{\mathrm{test}}_{\mathcal{T},n_{{\mathit{best},\mathcal{T}}},i}$ is the AUC value of task $\mathcal{T}$ of test data using a trained GCN with epoch $n_{{\mathit{best},\mathcal{T}}}$ in trial $i$. The division of the dataset at each trial $i$ is randomly performed each time.
EF (EF\_{}eval) is also obtained in the same way as in Eq.~(9).


\section{Results}
\subsection{Performance of Props. A and B}
The results of comparing the AUC of each dataset is shown in \tabref{table:auc_d} for the models of the Weave module, Prop. A, Prop. B, and Prop. A\&B.
Prop. A provided higher prediction performance than the Weave module in the MUV dataset but remained as accurate as the Weave module in HIV and PCBA. The accuracy of the Prop. B alone is almost the same as that of the Weave module, while the combination of the Prop. A and B yields a slightly higher accuracy. 

\begin{table}[tb]
	\caption{AUC of each dataset using Props. A and B. The value in bold-font is the best value for each model.}
	\label{table:auc_d}
	\centering
	\begin{tabular}{llrrrrr}
		\hline
		\multicolumn{1}{c}{\multirow{2}{*}{dataset}} & \multicolumn{1}{c}{\multirow{2}{*}{model}} & \multicolumn{5}{c}{distance}                                                                                          \\ \cline{3-7} 
		\multicolumn{1}{c}{}                         & \multicolumn{1}{c}{}                       & \multicolumn{1}{c}{1} & \multicolumn{1}{c}{2} & \multicolumn{1}{c}{3} & \multicolumn{1}{c}{4} & \multicolumn{1}{c}{5} \\ \hline
		\multirow{4}{*}{HIV}                         & Weave                                    & 0.796                 & 0.798                 & 0.795                 & 0.793                 & \textbf{0.801}        \\
		& Prop. A                                        & 0.796                 & \textbf{0.803}        & 0.799                 & 0.794                 & 0.798                 \\
		& Prop. B                                        & 0.794                 & 0.797                 & 0.797                 & 0.799                 & \textbf{0.806}        \\
		& Prop. A\&B                                      & \textbf{0.806}        & 0.798                 & 0.801                 & 0.800                 & 0.800                 \\ \hline
		\multirow{4}{*}{MUV}                         & Weave                                    & 0.680                 & 0.720                 & 0.739                 & 0.689                 & \textbf{0.743}        \\
		& Prop. A                                        & 0.706                 & \textbf{0.783}        & 0.735                 & 0.741                 & 0.754                 \\
		& Prop. B                                        & 0.723                 & \textbf{0.738}        & 0.714                 & 0.671                 & 0.736                 \\
		& Prop. A\&B                                      & 0.757                 & \textbf{0.760}        & 0.704                 & 0.737                 & 0.693                 \\ \hline
		\multirow{4}{*}{PCBA}                        & Weave                                    & 0.822                 & \textbf{0.824}        & 0.821                 & 0.821                 & 0.823                 \\
		& Prop. A                                        & 0.821                 & \textbf{0.825}        & 0.823                 & 0.823                 & 0.824                 \\
		& Prop. B                                        & 0.822                 & 0.821                 & 0.820                 & 0.822                 & \textbf{0.823}        \\
		& Prop. A\&B                                      & 0.819                 & 0.821                 & \textbf{0.823}        & 0.822                 & 0.821                 \\ \hline
	\end{tabular}
\end{table}

\begin{figure}[tb]
	\begin{center}
		\includegraphics[width=.95\linewidth]{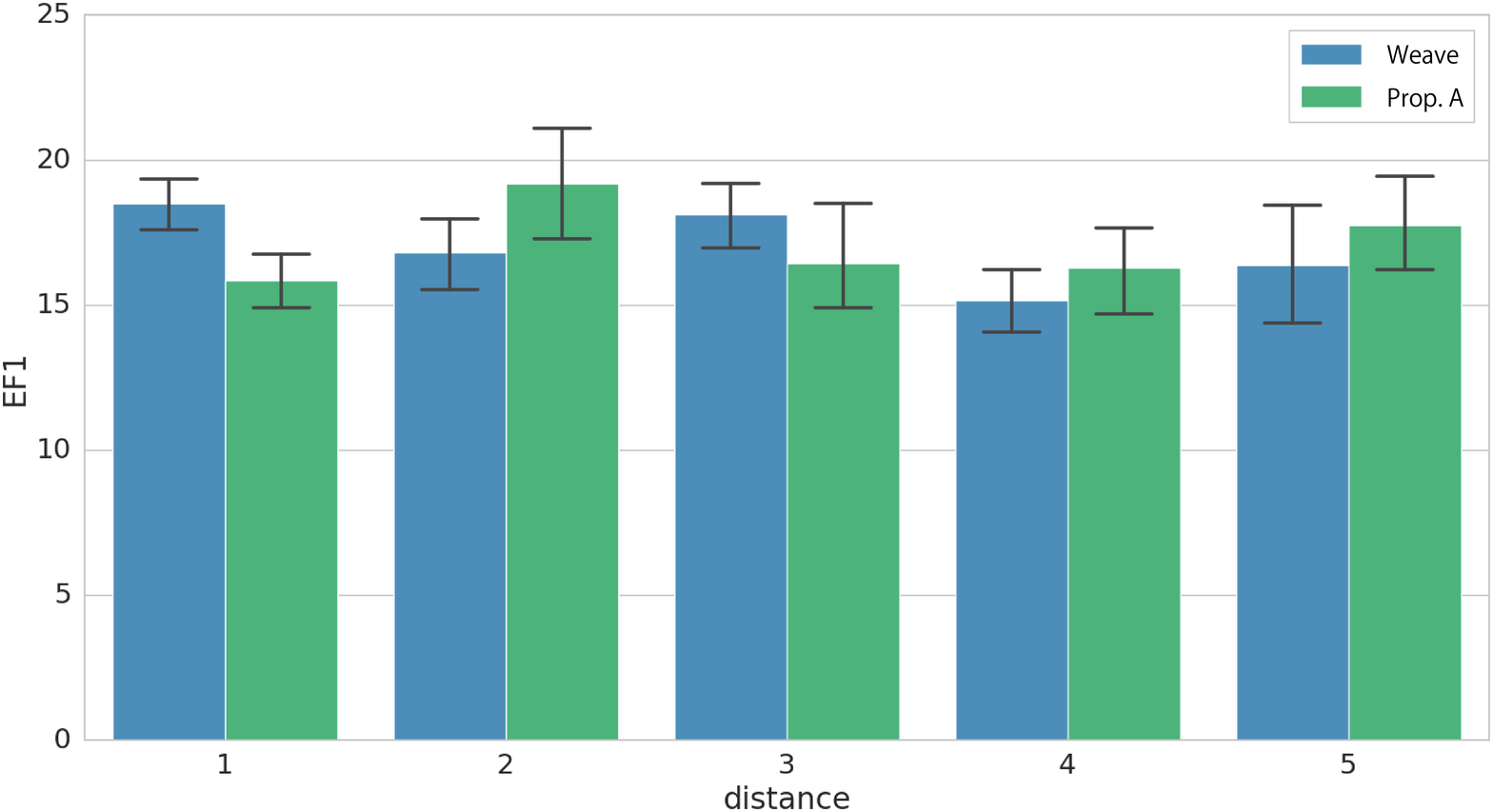}
		\caption{ EF$_{1\%}$ of HIV dataset using Prop. A}
		\label{fig:01_hiv_ef1}
	\end{center}
	\begin{center}
		\includegraphics[width=.95\linewidth]{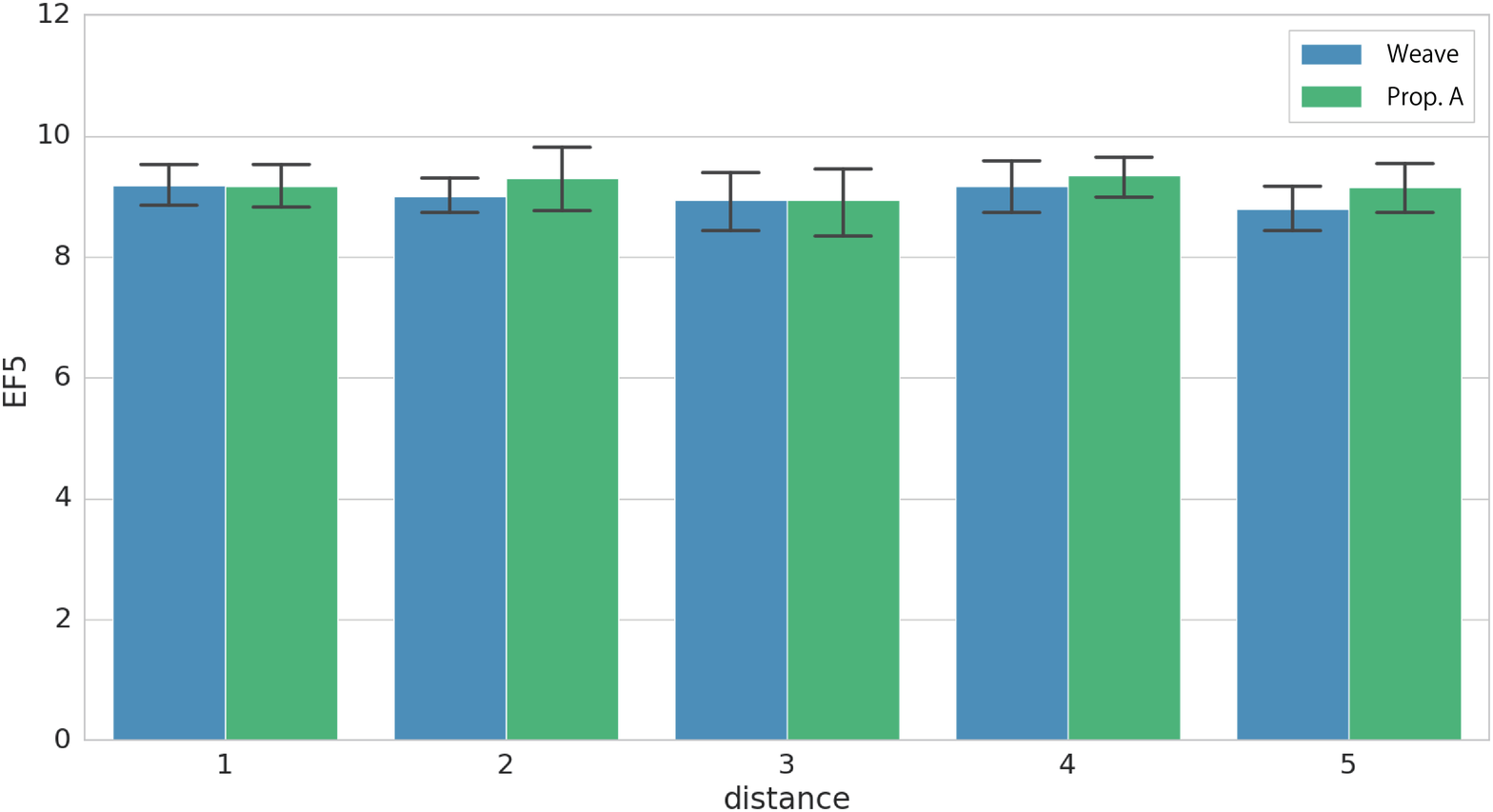}
		\caption{ EF$_{5\%}$ of HIV dataset using Prop. A}
		\label{fig:01_hiv_ef5}
	\end{center}
\end{figure}

The distribution of EF of Prop. A in the HIV dataset is shown in Figs. \ref{fig:01_hiv_ef1} and \ref{fig:01_hiv_ef5}.
EF$_{1\%}$ was the highest at 19.2 when $\mathrm{distance}=2$.
When $\mathrm{distance}=4,5$, EF$_{5\%}$ was also slightly higher. 
This suggests that the feature between the distant atoms could be reflected.
For the MUV dataset, no such results were seen for both EF$_{1\%}$ and EF$_{5\%}$, and the performance was almost the same as that of the Weave module.

\begin{figure}[tb]
	\begin{center}
		\includegraphics[width=.95\linewidth]{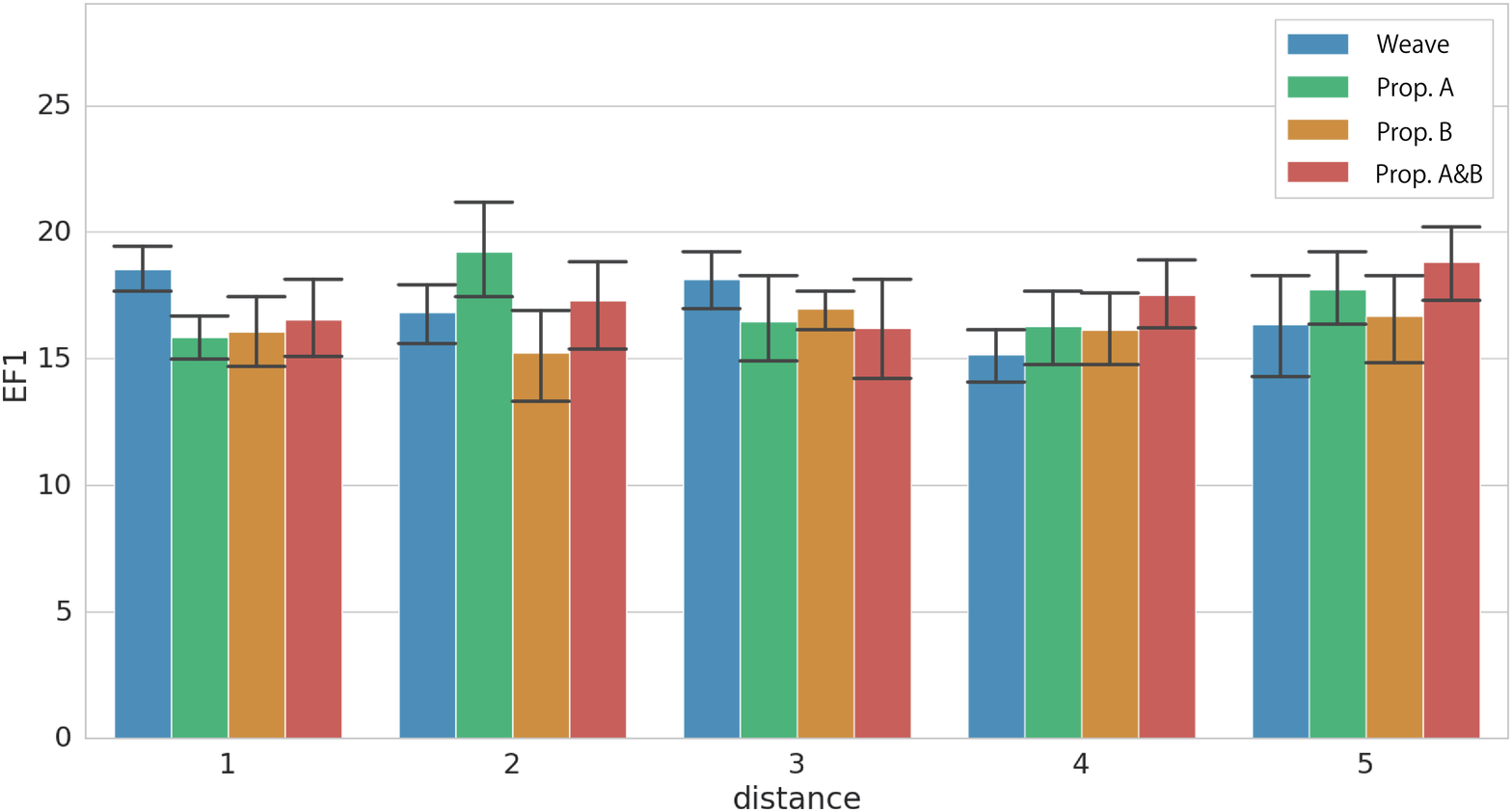}
		\caption{The EF$_{1\%}$ of HIV dataset using Prop. B}
		\label{fig:02_hiv_ef1}
	\end{center}
	\begin{center}
		\includegraphics[width=.95\linewidth]{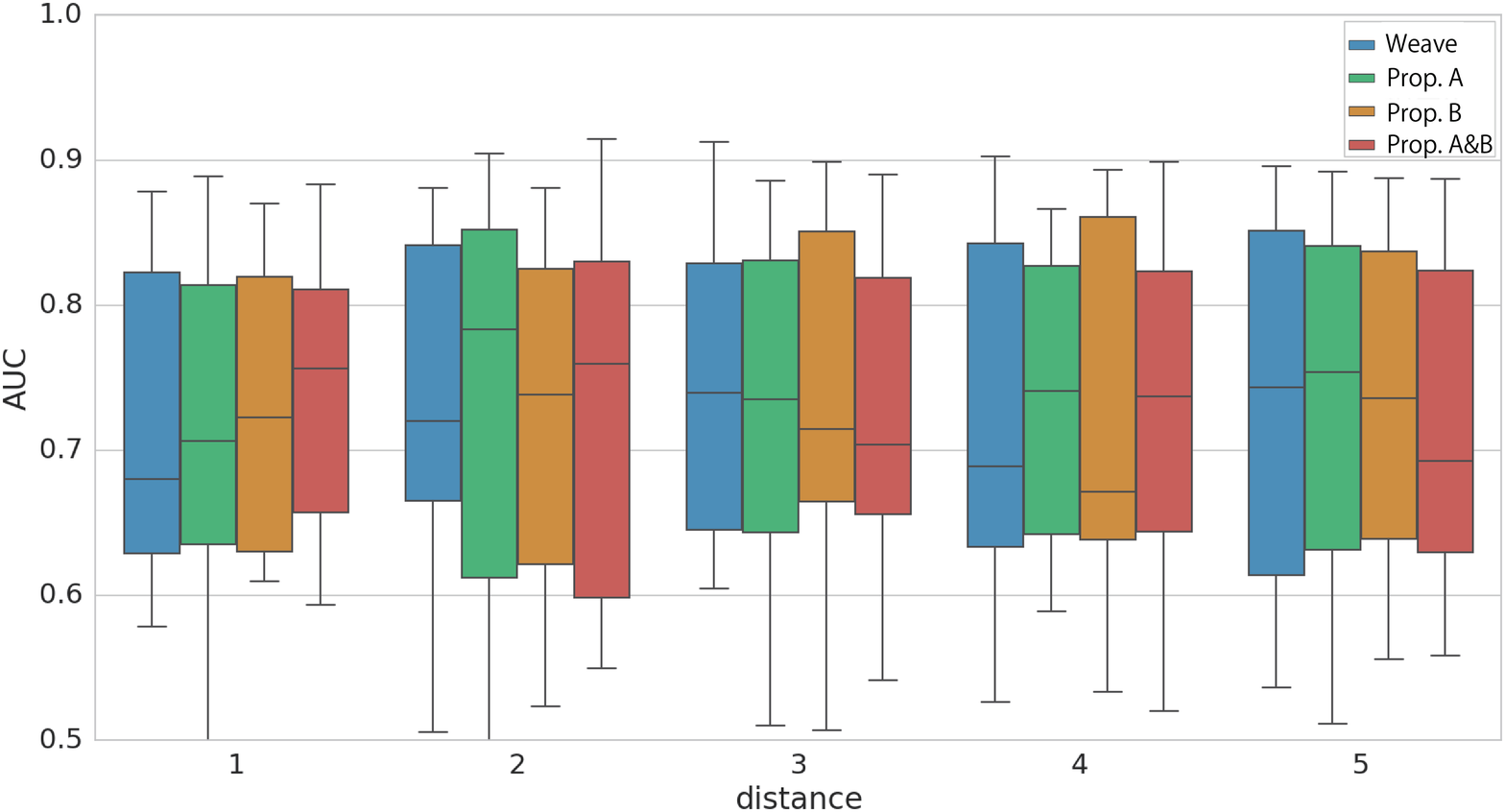}
		\caption{ AUC of MUV dataset using Prop. B}
		\label{fig:02_muv_auc}
	\end{center}
\end{figure}

Figs. \ref{fig:02_hiv_ef1} and \ref{fig:02_muv_auc} show the distributions of the AUC values. 
\figref{fig:02_muv_auc} shows the distribution by AUC for each task before median operation using Eq. (9).
The comparison of the model combining Prop. A and B with the Weave module and Prop. A shows that the variation of the AUCs on MUV tasks were decreased. The AUC value of Prop. A\&B was higher than others, in particular $\mathrm{distance} = 1, 2$ by separating the weights of nearby atoms.
In addition, the model combining Props. A and B obtained an EF$_{1\%}$ value of 18.8 when $\mathrm{distance}=5$. When $\mathrm{distance}=4, 5$, the performance was better compared to that of Prop. A.

\subsection{Results of Prop. C}
The prediction results for the HIV and MUV datasets are listed in \tabref{table:auc_gather} with respect to Prop. C, three functions, and the Weave module for assembling pair features.
The models of the linear and quadratic functions of Prop. C have a higher AUC value. The improvement by the model of the step function was not significant.

\begin{table}[tb]
	\caption{AUC of each dataset using Prop. C. The value in bold-font represents the best value for each model.}
	\label{table:auc_gather}
	\centering
	\begin{tabular}{llrrrrr}
		\hline
		\multicolumn{1}{c}{\multirow{2}{*}{dataset}} & \multicolumn{1}{c}{\multirow{2}{*}{model}} & \multicolumn{5}{c}{distance}                                                                                          \\ \cline{3-7} 
		\multicolumn{1}{c}{}                         & \multicolumn{1}{c}{}                       & \multicolumn{1}{c}{1} & \multicolumn{1}{c}{2} & \multicolumn{1}{c}{3} & \multicolumn{1}{c}{4} & \multicolumn{1}{c}{5} \\ \hline
		\multirow{4}{*}{HIV}     & Weave               & 0.796          & 0.798          & 0.795          & 0.793 & \textbf{0.801} \\
		& step               & 0.766          & 0.767          & 0.765          & 0.769 & \textbf{0.772} \\
		& linear               & 0.799          & 0.798          & 0.803          & 0.799 & \textbf{0.807} \\
		& quadratic               & 0.796          & 0.791          & \textbf{0.803} & 0.798 & 0.803          \\ \hline
		\multirow{4}{*}{MUV}     & Weave               & 0.680           & 0.720           & 0.739          & 0.689 & \textbf{0.743} \\
		& step               & 0.629          & \textbf{0.721} & 0.692          & 0.677 & 0.690           \\
		& linear               & 0.731          & \textbf{0.749} & 0.687          & 0.713 & 0.729          \\
		& quadratic               & \textbf{0.752} & 0.742          & 0.713          & 0.722 & 0.702          \\ \hline
	\end{tabular}
\end{table}

\section{Discussion}

\subsection{Effect of distance correction on molecular graph}
Because the number of ring structures was approximately three times the number of compounds in the dataset, it was useful to correct the distances on the graph for atoms in the ring structure using Prop. A.
In this study, we focused on the atoms in the ring structure, and considered the inter-atom distance $d$ in the ring structure as $\left\lceil d/2 \right\rceil$, not based on covalent bonds.
This unusual molecular graph structure allows us to modify the interatomic distance on the graph in the ring structure to correlate with the interatomic distance on the conformation.
Given the assumption that the learning of the ring structure by the GCN can be done by correcting the distance on the graph in a ring structure from the result of an experiment, it is better to omit the feature that an atom pair belongs to the same ring.
Furthermore, different rings with different interatomic bonds, such as benzene and cyclohexane, cannot be distinguished by GCNs, including the Weave module and the proposed method. The prediction performance can be expected to improve by considering the type of bond in the ring.

\subsection{Transition of weight matrix norm}
\begin{figure}[tb]
	\begin{center}
		\includegraphics[width=.95\linewidth]{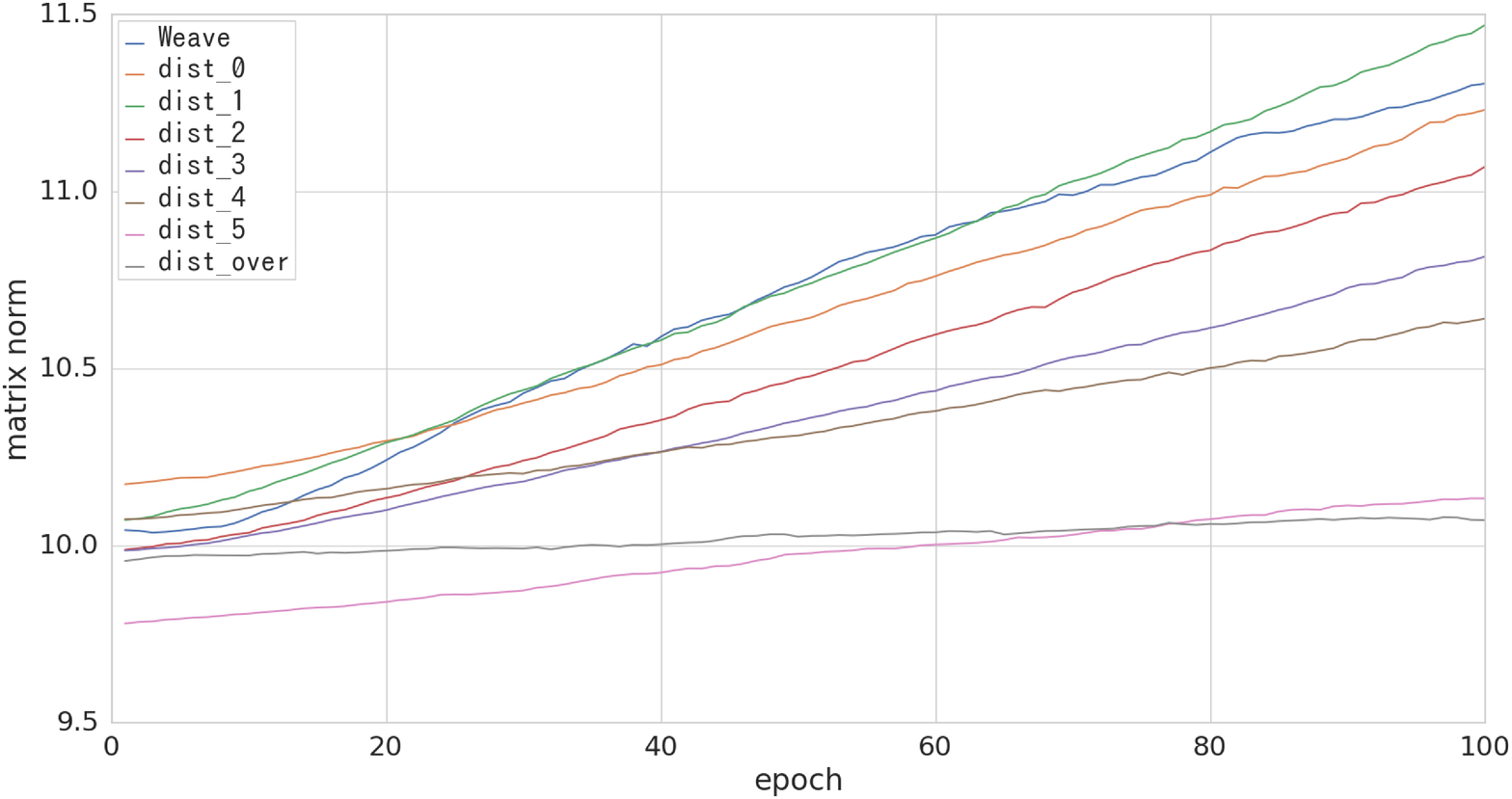}
		\caption{Comparison of weight matrix norms ($\bm{W}_{\mathit{PA}}^0$ and $\bm{W}_{\mathit{PA}_{\mathit{dist}_n}}^0$) in HIV dataset}
		\label{fig:hiv_dp23f}
	\end{center}
	\begin{center}
		\includegraphics[width=.95\linewidth]{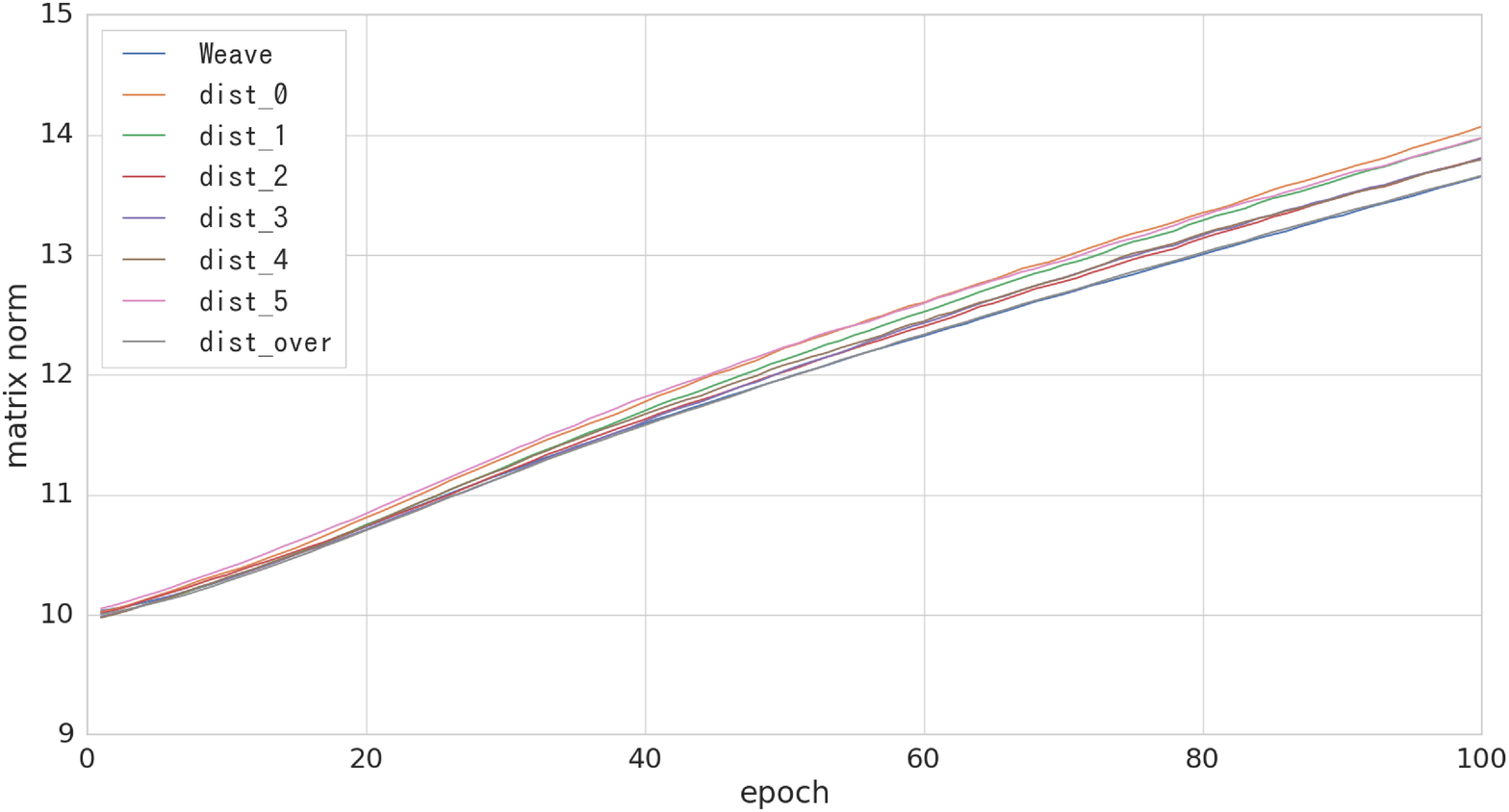}
		\caption{Comparison of weight matrix norms ($\bm{W}_{\mathit{PA}}^1$ and $\bm{W}_{\mathit{PA}_{\mathit{dist}_n}}^1$) in HIV dataset}
		\label{fig:hiv_dp23s}
	\end{center}
\end{figure}

For the Weave module and Prop. B, we examined how the weighting matrix changed as the learning progressed.
The Frobenius norm $\|\bm{W}\|_F = \sqrt{\sum_{i,j}w^2_{ij}}$ of each weight matrix $\bm{W}=(w_{ij})$ at the HIV dataset and maximum atomic pair distance 5 was determined.
The 0th and 1st layers of the Weave module are as shown in \figref{fig:hiv_dp23f} and \figref{fig:hiv_dp23s}, respectively.
dist\_$n$ in each figure is the norm of the weight matrix when the distance is $n$, and dist\_over is the norm of the weight matrix when the maximum atomic pair distance is greater than 5.

According to \figref{fig:hiv_dp23f}, dist\_0, dist\_1, and dist\_2 have roughly equal slopes compared to the Weave module; however, dist\_5 and dist\_over have gentle slopes. In the 0th layer of the Weave module, distant atom pairs were not considered very important because the value of the weight matrix does not fluctuate excessively, and they are incorporated with more emphasis on the close atom pairs. Accordingly, it is possible to improve the model performance by using different weights for the atom pair distance 0 to 2 and the other atom pairs.
On the other hand, in \figref{fig:hiv_dp23s}, the slopes of the Weave module and dist\_$n$ were approximately equal. The values of the weight matrix were found to fluctuate significantly in the 1st layer of the Weave module, thus emphasizing not only pair features with close interatomic distances but also those with larger interatomic distances.
Therefore, it may not be necessary to divide the weight matrix for each distance in the 1st layer, and it may be effective to change the composition of the weight matrix for each layer of the Weave module.

\section{Conclusion}
In this study, three types of improvements were made to the operation of converting a pair feature to an atom feature in the Weave module~\cite{weave}, which is one of the GCN models of compounds. \\
{\it A. correction of the distance on the graph in the ring structure in the compound}: By changing distance $d$ of the atom pair contained in the ring structure to $\left\lceil d/2 \right\rceil$, the distance on the graph was corrected to correlate with the distance on conformation. As a result of the evaluation experiment, the prediction accuracy is improved compared to the Weave module, and features between distant atoms were also successfully used.\\
{\it B. convolution of paired features with different weights for different distances on the graph}:
We attempted to generalize the model by using different weights for each distance in the convolution process combined with the proposed correction of the distance on the graph in the ring structure in the compound. The prediction accuracy was higher when performing convolution with different weights for each distance compared to the Weave module. According to the analysis of the weight matrix dynamics, the proposed method was found to be useful, especially in the 0th layer of the Weave module.\\
{\it C. Assembly of paired features based on distances on the graph}: We proposed a method of incorporating pair features that emphasize the atoms in the vicinity of the atom of interest by using coefficients according to the distance. We achieved some improvement in the prediction accuracy by assembling paired features by using linear and quadratic weights.

We intend to work on the following topics in the future:
\begin{itemize}
\item Distinguishing ring types based on atomic bonds may improve prediction accuracy.
\item Because Weave-module vector transformation operation is complicated, it may not be possible to achieve a significant improvement in the accuracy simply by improving the transformation operation from pair features to atom features. In other transformation operations, it may be necessary to make improvements by utilizing distance features.
\item It is worthwhile to verify that this improvement is also effective for other tasks of compound supervised learning, e.g., drug-like compound filter~\cite{qex}, side-effect prediction \cite{sideeffect}, toxicity prediction~\cite{tox}, and stability prediction~\cite{ppb,ppb2}.
\end{itemize}

\section*{Acknowledgment}
This work was partially supported by 
KAKENHI (Grant No. 17H01814, 17J06897, and 18K18149) from the Japan Society for the Promotion of Science (JSPS),
the Core Research for Evolutional Science and Technology (CREST) ``Extreme Big Data'' (Grant No. JPMJCR1303) from the Japan Science and Technology Agency (JST),
the Program for Building Regional Innovation Ecosystems ``Program to Industrialize an Innovative Middle Molecule Drug Discovery Flow through Fusion of Computational Drug Design and Chemical Synthesis Technology'' from the Japanese Ministry of Education, Culture, Sports, Science and Technology (MEXT),
the Research Complex Program ``Wellbeing Research Campus: Creating new values through technological and social innovation'' from JST,
and the Platform Project for Supporting Drug Discovery and Life Science Research (Basis for Supporting Innovative Drug Discovery and Life Science Research (BINDS)) from the Japan Agency for Medical Research and Development (AMED) (Grant No. JP18am0101112).

\end{document}